\documentclass[12pt]{article}
\usepackage[utf8]{inputenc}
\usepackage{array, tabularx, booktabs}
\usepackage{braket}
\usepackage[english]{babel}
\usepackage{jheppub}
\usepackage{amsmath} \usepackage{amsfonts} \usepackage{amssymb}
\usepackage{wasysym}
\usepackage[dvipsnames]{xcolor}
\usepackage{graphicx}
\usepackage{multirow}
\usepackage{here}
\usepackage{epsfig}
\usepackage{epstopdf}
\usepackage{soul}
\usepackage{hyperref}
\usepackage{float}
\usepackage{xcolor}
\usepackage{graphicx}
\usepackage{multirow}
\usepackage{ulem}
\usepackage{array}
\usepackage{cancel}
\newcolumntype{L}{>{\displaystyle}l}
\newcolumntype{C}{>{\displaystyle}c}
\newcolumntype{R}{>{\displaystyle}r}

\newcommand{\be}{\begin{equation}}
\newcommand{\ee}{\end{equation}}
\newcommand{\bea}{\begin{eqnarray}}
\newcommand{\eea}{\end{eqnarray}}

\title{Islands in the Fluid: Islands are Common in Cosmology}
\author[a,b,c]{Ido Ben-Dayan}
\author[d,f]{Merav Hadad}
\author[c,e]{Elizabeth Wildenhain}
\affiliation[a]{Physics Department, Ariel University, Ariel 40700, Israel}
\affiliation[b]{Berkeley Center for Cosmological Physics, University of California, Berkeley, CA 94720, USA}
\affiliation[c]{Department of Physics, University of California, Berkeley, CA 94720, USA}
\affiliation[d]{Department of Natural Sciences, The Open University of Israel, Raanana 43107, Israel}
\affiliation[e]{Center for Theoretical Physics,\\
University of California, Berkeley, CA 94720, U.S.A.}
\affiliation[f]{Astrophysics Research Center of the Open university (ARCO), The Open University of Israel, P.O Box 808, Ra’anana 4353701, Israel}
\emailAdd{ido.bendayan@gmail.com}
\emailAdd{meravha@openu.ac.il}
\emailAdd{elizabeth\_wildenhain@berkeley.edu}

\abstract{We discuss the possibility of entanglement islands in cosmological spacetimes with a general perfect fluid with an equation of state $w$. We find that flat universes with time-symmetric slices where the Hubble parameter vanishes always have islands on that slice. We then move away from such slices, considering still universes with a general perfect fluid. Under the local thermal equilibrium assumption, the comoving entropy density $s_c$ is constant. As a result, the conditions for an island become an inequality between the energy density (or Hubble parameter) and the temperature at some time of normalization. The consequences are that islands can exist for practically all fluids that are not radiation, \textit{i.e.} $w\neq 1/3$. We also discuss the ramifications of our results for universes with spatial curvature. Finally, we show that islands occur in the Simple Harmonic Universe model which has no classical singularity at the background level, in contrast to all previous examples where islands occurred only in space-times with singularities.}

\begin{document}

\maketitle

\section{Introduction}
There has been much recent attention given to the Quantum Extremal Surface (QES) prescription~\cite{Ryu:2006bv, RyuTak06b, Hubeny:2007xt, Faulkner:2013ana, Engelhardt:2014gca} as a means of investigating gravitational systems. Originally derived in the context of AdS/CFT, the QES prescription is a method for relating entropies in the boundary CFT to quantities in the bulk spacetime. It was later shown that the prescription also follows from a gravitational path integral, and it is becoming apparent that the prescription contains deeper physics than it originally seemed~\cite{Bousso:2022hlz}. 

The QES prescription can be used to derive nontrivial features of quantum gravity, a notable example being the calculation of the Page curve for Hawking radiation of an evaporating black hole~\cite{Penington:2019npb, Almheiri:2019psf}. The key to this calculation is the contribution of a region of the bulk spacetime disconnected from the boundary region of interest to the QES formula. Such a region has come to be called an \textit{entanglement island}.

As pointed out in~\cite{Hartman:2020khs, Bousso:2022gth, Espindola:2022fqb}, it is of interest to see if this newly recognized feature of the QES prescription, namely, the involvement of islands, can teach us anything about Cosmology, where we have a poor understanding of quantum gravity. Toward this end, Ref.~\cite{Hartman:2020khs} investigated when it was possible for a region in a flat FLRW cosmology to be an island for a region in a nongravitating auxiliary system. Ref.~\cite{Hartman:2020khs} derived three necessary conditions for a region to be an island and found that all conditions could be satisfied in flat universes with a negative cosmological constant. Refs.~\cite{Bousso:2022gth, Espindola:2022fqb} expanded this analysis to FLRW cosmologies with nonzero curvature and derived an additional necessary condition for subsets of closed universes to be islands. The entire Cauchy slice of a closed universe was always found to satisfy all necessary conditions. Subsets of closed and open universes with negative cosmological constants were also found to satisfy the conditions in certain parameter regimes. These viable island candidates were found on or near time-symmetric slices (where the scale factor $a$ satisfies $\dot{a}=0$).

All these works assumed the matter entropy in the cosmology is generated by radiation, modeled by entangling the cosmological spacetime with the reference spacetime in a thermofield-double-like state. There are, however, interesting cosmological models that involve more general fluids \cite{Kolb:1990vq}, including simple cosmologies without singularities \cite{Graham:2011nb}. In this work, we wish to extend the analysis of \cite{Hartman:2020khs, Bousso:2022gth, Espindola:2022fqb} by relaxing the assumption that the contents of the cosmology is radiation. We will thus search for islands in FLRW cosmologies with a general fluid with a constant equation of state fulfilling the Null Energy Condition $w\geq-1$. 

\subsection{Outline and Summary}
The main objective of this work is to generalize the analysis of \cite{Hartman:2020khs, Bousso:2022gth, Espindola:2022fqb} to cosmologies containing general fluids. In section \ref{section:preliminaries}, we review the Quantume Extremal Surface prescription and the necessary conditions for islands derived in \cite{Hartman:2020khs, Bousso:2022gth, Espindola:2022fqb}. We also describe a specific model of which we will make use, that of a gravitating universe and a nongravitating universe entangled in a thermofield-double-like state.

As a prelude to applying the conditions to classes of spacetimes, section \ref{section:applyingConditions} reviews the relevant cosmological thermodynamics, including the argument from \cite{Kolb:1990vq} that the comoving entropy density is a constant in time under the assumption of local thermal equilibrium.
 
In section \ref{section:conditionsRearranged}, we write out the necessary conditions for islands in full generality, with a general fluid. The crucial property of constant comoving entropy density allows us to phrase the necessary conditions for islands for any FLRW universe as inequalities regarding the comoving entropy density. This constant comoving entropy density $s_c$ has to be larger than a time-dependent term $s_c\geq a^2/4G_{N} f(\eta,\chi)$, where $a$ is the scale factor, $G_{N}$ Netwon's constant, $\chi$ is a generalized radial coordinate, $\eta$ is a conformal time coordinate, and $f(\eta,\chi)$ is a fudge factor that is different between the different conditions and depends on geometrical quantities such as the Hubble parameter, the comoving volume, and the comoving area. In many cases, for a given choice of $\chi$, fulfilling one inequality will imply the fulfillment of all others, making it what we call an ``encompassing" condition for a region to be a viable island candidate.

Since time-symmetric slices have frequently been found to contain viable island candidates, we start with considering such slices in \ref{section:TSS}. We arrive at the conclusion that a time-symmetric slice in a flat universe always contains an island. In open universes a time-symmetric slice contains an island if $s_c>a_{ts}^2/2G_{N}$, where $a_{ts}$ is the scale factor at the turnaround time. In closed universes, one cannot combine the conditions into a single encompassing inequality. However, for $\chi\simeq \pi/2$ on a time-symmetric slice in a closed universe, one does get a single condition of $s_c>a_{ts}^2/\pi G_{N}$.

The existence of time-symmetric slice could occur with rather unconventional energy components, e.g. \cite{Artymowski:2019cdg}. To make contact with models usually considered in Cosmology, we distinguish a subset of universes with ``conventional" components. These are universes that can only include any number perfect fluids, with positive definite energy density, a CC and spatial curvature. We elaborate on the existence of islands in this conventional subset in section \ref{section:TSS}.

Section \ref{section:flat} applies the conditions to flat universes with a general perfect fluid away from time-symmetric slices. We address two separate questions. \textbf{1)} Assuming that we are allowed to consider any (subplanckian) energy density $\rho_0$ and temperature $T_0$ of a perfect fluid $w$, at some normalization time $t_0$ is there an island? and \textbf{2)} Assuming that we start with some given $\rho_0$, $T_0$, and $w$ at $t_0$, will islands form when we evolve forwards or backwards in time?

In answer to the first question, we find that all conditions can be simultaneously satisfied if the island is large enough, and a certain inequality is fulfilled between the energy density $\rho_0$ and the temperature $T_0$ at some time of normalization $t_0$, $G_{N} \rho_0>\mathcal{O}(1) \max \left\{T_0^2, \frac{T_0}{\chi_{0,phys.}}\right\}$. However, it does seem that for the simple case of radiation the inequalities are not fulfilled. This contrasts the results of \cite{Hartman:2020khs, Bousso:2022gth, Espindola:2022fqb} in which all viable island candidates appeared on or near time-symmetric slices. 

The answer to the second question is related to the Dominant Energy Condition (DEC) $|w|\leq 1$. We find that islands will always exist when we evolve forward in time if the DEC is violated $w\geq 1$ for a large enough spatial region $\chi \gg 1$. The converse is true for backward time evolution. For $\sqrt{8\pi G_{N}} \ll t\ll t_0$ islands form away from the planckian regime, if the DEC, $|w|\leq 1$, $\chi \gg1 $, and a certain inequality regarding the temperature $T_0$ are fulfilled, $\frac{1}{2\pi t_0}\left(\frac{t_0}{t}\right)^{(1-w)/(1+w)}\geq T_0\geq  \frac{1}{2\pi t_0}$. We demonstrate an island explicitly with non-relativistic dust $w=0$. Again, radiation is a very special case that does not produce an island when taking into account all the parameters. 

The reason that radiation is special is due to the conformal nature of radiation. For radiation, there is only a single energy scale which is the temperature $T_0$, and the conditions are not satisfied. For all other fluids, there are other energy scales such as $\rho_0$ or the mass of the dust particle $m$. As a result, there is a hierarchy between these energy scales and the conditions for islands can be fulfilled. In a certain sense, this argument explains the results of \cite{Hartman:2020khs, Bousso:2022gth, Espindola:2022fqb}, as spatial curvature and/or a CC introduced another energy scale that enabled the existence of a time-symmetric slice where conditions for islands are favorable.

In sections \ref{section:closedandopen}, we specialize the conditions to closed and open universes with a general fluid. In the regime of fluid domination for any $w$, if the scale factor is monotonically growing, islands can occur only at the time of normalization $t_0$, before that time, and up to some finite time later. In this section, we also consider a specific example of interest, which is a closed universe  without singularities: the ``Simple Harmonic Universe." This model which contains a fluid with $w=-2/3$, has a negative CC and positive spatial curvature $\Lambda<0$, $k=+1$. Its scale factor is periodic, and we find that all island conditions can be simultaneously satisfied near slices on which the scale factor reaches a minimum. Furthermore, entire Cauchy slices of this universe are islands, which follows directly from the results of~\cite{Bousso:2022gth}. This is the first example of a possible island in space-time without singularities. 

In brief, islands in Cosmology require either a time-symmetric slice or a condition on the temperature and energy density of the fluid, and they are not necessarily accompanied by a singularity at least at the background level. Islands are common in Cosmology!

\section{Preliminaries}
\label{section:preliminaries}
In this section, we review the Quantum Extremal Surface (QES) prescription, the island rule, and the necessary conditions for islands derived in Refs.~\cite{Hartman:2020khs, Bousso:2022gth, Espindola:2022fqb}. We also review FLRW cosmologies and the relevant thermodynamics.

\subsection{The Quantum Extremal Surface Prescription}
The QES prescription \cite{Engelhardt:2014gca, Faulkner:2013ana, Hubeny:2007xt, Ryu:2006bv, RyuTak06b} was originally discovered as a way to compute entropies in AdS/CFT, although evidence is growing that it is a much deeper statement. The prescription is, however, most simply stated in its AdS/CFT form. Given a region $R$ in a nongravitating boundary system, the QES prescription computes the entropy $R$ in terms of quantities in a dual bulk spacetime with semiclassical gravity:
\be
    S(\textbf{R})=S_{\mathrm{gen}}[EW(R)]~,
\ee
where $EW(R)$ is a bulk region called the entanglement wedge, $S_{\mathrm{gen}}(X)$ is the generalized entropy, the $S(\textbf{R})$ with bold-face \textbf{R} signifies the entropy computed by the QES prescription, and the use of a non-bold face R indicates the appearance of von Neumann entropy computed directly from the semiclassical state. The generalized entropy for bulk region $X$ is defined as
\begin{equation}
    S_{\mathrm{gen}}(X) = \frac{A(\partial X)}{4G_N}+S\left(X\right)~,
\end{equation}
where $A(\partial X)$ is the area of the boundary of $X$, $G_{N}$ is Newton's constant, and $S(X)$ is the von Neumann entropy of the density operator of the quantum field theory state reduced to $X$. The entanglement wedge for boundary region $R$ is a bulk region that satisfies the following constraints:
\begin{enumerate}
    \item \textit{Homology}.~$EW(R)$ is bounded by a surface $\gamma$ in the bulk that is homologous to $R$.
    \item \textit{Stationarity}.~The generalized entropy of $EW(R)$ is stationary with respect to variations in $\gamma$.
    \item \textit{Minimality}.~Among all regions that satisfy the first two constraints, $EW(R)$ is the one that minimizes the generalized entropy.
\end{enumerate}
In the case where $R \subset \Sigma_R$ is a partial Cauchy surface in a nongravitating spacetime $M_R$ distinct from $M$, then the $EW$ reduces to the so-called ``island rule:"
\begin{enumerate}
    \item \textit{Homology}.~$EW(R) = I\cup R$, where $I\subset \Sigma_M$ and $I$ is compact.\footnote{More precisely, in the conformally compactified spacetime, the boundary of the image of $I$ does not intersect with the conformal boundary of $M$.}
    \item \textit{Stationarity}.~$S_{\mathrm{gen}}(I\cup R)$ is stationary with respect to variations of its boundary $\partial I$.
    \item \textit{Minimality}.~Among all regions that satisfy the first two constraints, the choice of $I$ minimizes $S_{\mathrm{gen}}(I\cup R)$.
\end{enumerate}
 It is this form of the QES prescription that concerns this work, as we are interested when a non-empty island ($I$) can occur.

\subsection{Necessary Conditions for Islands}
Refs.~\cite{Hartman:2020khs, Bousso:2022gth} derived a set of four conditions necessary for the existence of a nonempty island in a spacetime $M$ entangled with a reference system $M_R$. We will refer to a region that satisfies these four conditions as a ``viable island candidate," while a region for which we consider the four conditions but have not yet checked them is an ``island candidate." Assuming the global quantum state on Cauchy slices $\Sigma_M \cup \Sigma_R$ is pure, the four conditions are:
\begin{enumerate}
\item $S(I)>\frac{A(\partial I)}{4G_N}$.
\item $I$ is quantum normal.
\item $G$ is quantum normal.
\item For spatially closed $M$ and $I \neq \subset \Sigma_M$, $S(G)>\frac{A(\partial I)}{4G_N}$.
\end{enumerate}
\begin{figure*}
    \centering
    \includegraphics[width=.8\textwidth]{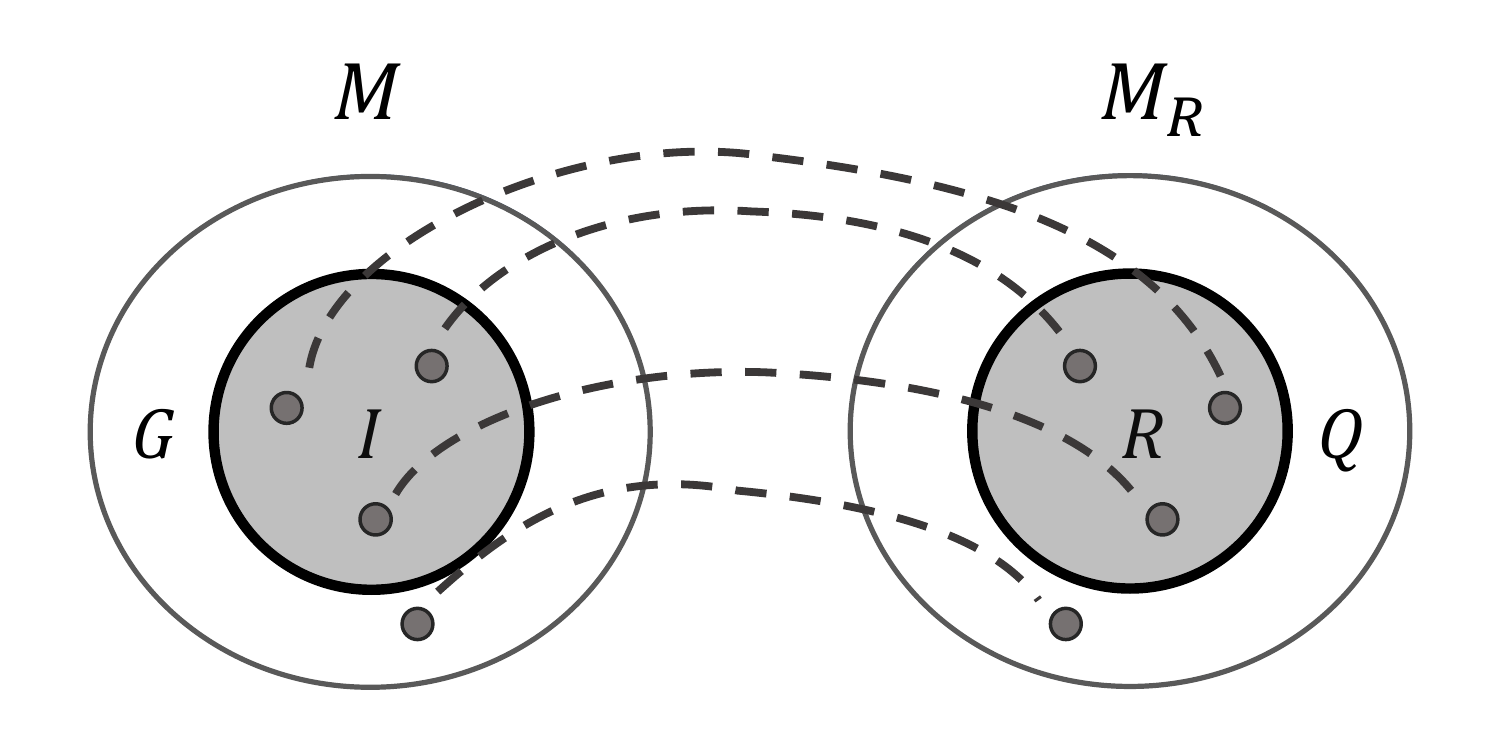}
    \caption{Schematic figure of $M$ and $M_R$. A region in $M$ can be an island ($I$) for some reference region $R\subset M_R$ only if it satisfies the four necessary conditions. $G$ is the complement of $I$, and $Q$ is the complement of $R$. Condition 1 requires a large degree of entanglement, represented by the dashed lines. }
    \label{fig:schematicM}
\end{figure*}
Here $G$ is the complement of $I$, namely $G\equiv \Sigma_M\backslash I $, and $Q$ is the complement of $R$: $Q\equiv \Sigma_R\backslash R$, see Figure \ref{fig:schematicM}. Condition 1 comes from the requirement that an island must be ``worth" its cost in area, given that the QES prescriptions requires minimization of the generalized entropy. Conditions 2 and 3 are consequences of the fact that $I\cup R$ and $G \cup Q$ are quantum extremal. Island candidates that are subsets of a spatially closed universe must satisfy condition 4 in order to win out over the entire Cauchy slice, which is always a viable island candidate. See Refs.~\cite{Hartman:2020khs, Bousso:2022gth} for a full derivation.

Ref.~\cite{Hartman:2020khs} used these conditions to search for islands in flat FLRW spacetimes containing radiation entangled with a purifying reference system. Refs.~\cite{Bousso:2022gth, Espindola:2022fqb} generalized this analysis to FLRW spacetimes with non-zero curvature. In this work, we dispense with the assumption that the contents of the universe are radiation, instead considering an arbitrary perfect fluid.

\subsection{Thermofield-doubled FLRW Construction}

In order for the QES prescription to force an island into the entanglement wedge of some space-time region, there must be entanglement between that candidate island and the reference region. That way, including the island in the entanglement wedge purifies some of the matter in the reference region, and if that decrease in matter entropy makes up for the area cost of the island, the island will be in the entanglement wedge. It is unclear in Cosmology what could be the source of this entanglement with an external system. If we wish to make statements about whether a viable island candidate is in fact an island for some reference region, we require a specific model of that entanglement. In the spirit of Refs. \cite{Hartman:2020khs, Bousso:2022gth, Espindola:2022fqb}, we will consider a construction that essentially puts in a high degree of entanglement by hand. This construction consists of a gravitating FLRW spacetime, purified by a reference spacetime $M_R$ conformally related to the gravitating spacetime. There is freedom in the choice of $M_R$; following Ref. \cite{Bousso:2022gth} we select
\begin{equation}
    ds_R^2/\ell^2 = -d\eta_R^2 + d\chi_R^2 + f^2(\chi_R)d\Omega_R^2~,
\end{equation}
where $\ell$ is an arbitrary fixed length scale. In this construction, the gravitating FLRW spacetime $M$ and the reference spacetime $M_R$ are placed in a thermofield-double-like state, which is first constructed using two copies of $M_R$
\begin{equation}
    |\mathrm{TFD}\rangle\propto\sum_n e^{-\beta E_n}|n\rangle^*_1|n\rangle_2~.
\end{equation}
A conformal transformation by the scale factor $a^2$ is then applied to one of the copies to transform it to $M.$ In this work, we will assume the universe contains a general fluid rather than radiation; thus $\beta=\ell/(aT)$, where $T$ is the temperature of this fluid in $M$ at scale factor $a$. In this construction, the matter entropy in $M$ is a uniform thermal entropy density, in accord with the assumption of a perfect fluid in local thermal equilibrium.

\subsection{FLRW and General Perfect Fluids}
\label{section:applyingConditions}
The cosmological models we will consider are FLRW universes filled with general perfect fluid(s) with an equation of state $w\geq-1$. The metric for an FLRW universe is 
\be
ds^2=-dt^2+a(t)^2\left[d\chi^2+f^2(\chi)d\Omega^2\right]=a^2(\eta)\left[-d\eta^2+d\chi^2+f^2(\chi)d\Omega^2\right]~,
\ee
where $t$ denotes cosmic time and $\eta$ conformal time, which are related by $dt=a(\eta)d\eta$. We shall use both of them throughout the paper depending on convenience.  The factor $f(\chi)$ is $\sinh(\chi),\,\chi,\, \sin(\chi)$ for open, flat, and closed universe respectively. Notice that in contrast to some conventions, $\chi$ and $\eta$ are dimensionless while the scale factor $a$ has dimensions of length or time. For a general constant equation of state, $w_i$, the scale factor, and the energy density of fluid $i$ behave as
\be \label{eq:rhoi}
\rho_i=\rho_{0i}\left(\frac{a}{a_0}\right)^{-3(1+w_i)}~,
\ee
where the subscript ``$0$" denotes some time of normalization. Throughout its evolution, the universe is dominated by a single fluid except for transient periods.
In this parametrization, the Big Bang singularity is at $a=0$.\footnote{An exception is if the NEC is violated with temporary $w<-1$, which occur for example in some quintessence models or bouncing models, e.g. \cite{Ben-Dayan:2016iks, Ben-Dayan:2018ksd, Artymowski:2019cdg, Artymowski:2020pci}.} We are interested in islands away from the singularity where semiclassical analysis can be trusted.
The governing equations of motion are given by the first Friedmann equation and the continuity equations that hold separately for each fluid:
\bea
H^2=\sum_i\frac{8\pi G_{N}}{3}\rho_{i}-\frac{k}{a^2}\pm\frac{1}{t_{\Lambda}^2}~, \label{eq:Friedmann1}\\
\frac{d\rho_i}{dt}=-3H(\rho_i+p_i)~,\label{eq:continuity_i}
\eea
where $t_{\Lambda}=\sqrt{3/|\Lambda|}$ and $k=\pm1,0$ is the usual spatial curvature.
The entropy density of a fluid $i$ is given by
\be \label{eq:sth}
s_i=\frac{\rho_i+p_i-\mu_i n_i}{T_i}~,
\ee
where $\rho_i$ is the energy density, $p_i$ the pressure, $n_i$ the number density, $\mu_i$ the chemical potential, and $T_i$ the temperature. Neglecting the chemical potential $\mu_i$, we have
\be
\label{eq:sthw}
s_i=\frac{(1+w_i)\rho_{0i}}{T_i}\left(\frac{a}{a_0}\right)^{-3(1+w_i)}~,
\ee
where we have used the equation of state of the i-th fluid $p_i=w_i\rho_i$. In local thermal, equilibrium the total comoving entropy is conserved:
\be
s_{c}=s\,a^3=const.
\ee
If the fluids are decoupled such a conservation occurs for every fluid separately assuming it is in thermal equilibrium with itself. The proof of this (taken from \cite{Kolb:1990vq}) is as follows. Applying the laws of thermodynamics to a comoving volume element yields
\begin{equation}
    \label{eq:secondlaw}
    TdS = d(\rho V) + pdV = d[(\rho+p)V]-Vdp~,
\end{equation}
where $V$ is the physical volume ($a^3\mathcal{V}(\chi)$), $\rho$ is the equilibrium energy density, $p$ is the equilibrium pressure, and $S$ is the entropy per comoving volume. The energy density and pressure are related via
\begin{equation}
    \frac{\partial^2S}{\partial T \partial V}=\frac{\partial^2 S}{\partial V\partial T}~,
\end{equation}
which implies,
\begin{equation}
    \label{eq:comovingproof1}
    dp=\frac{\rho+p}{T}dT~.
\end{equation}
Substituting into Eq. \ref{eq:secondlaw}, implies that
\begin{equation}
    dS = \frac{1}{T}d[(\rho+p)V]-(\rho+p)V\frac{dT}{T^2} =d\left[\frac{(\rho+p)V}{T}+\mathrm{const}\right]~
\end{equation}
and thus that
\begin{equation}
\label{eq:volume_entropy}
    S = \frac{a^3 (\rho+p)}{T}
\end{equation}
up to a constant. The first law of thermodynamics can be expressed as
\begin{equation}
    d[(\rho+p)V]=Vdp~,
\end{equation}
which when substituted into Eq. \ref{eq:comovingproof1} yields
\begin{equation}
    d\left[\frac{(\rho+p)V}{T}\right]=0~.
\end{equation}
Therefore, the comoving entropy is conserved in local thermal equilibrium.

In this work we will consider a single fluid, deferring more realistic analysis to future work. 
Because the total $s_c$ is constant, and assuming a single fluid in thermal equilibrium, the temperature of the fluid will redshift as $T\sim a^{-3w}$, which imposes the relation $s_{th}\sim\rho^{1/(1+w)}$.
Using \eqref{eq:comovingproof1}, \eqref{eq:volume_entropy}, and $p=w\rho$, one gets
\bea
\rho=\rho_0 \left(\frac{T}{T_0}\right)^{(1+w)/w}~,\\
\quad s_{th}=(1+w)\frac{\rho_0}{T_0}\left(\frac{T}{T_0}\right)^{1/w}~.\eea  
Hence, at the time of normalization $t_0$,
\be \label{eq:sc}
s_c=(1+w)\frac{\rho_0}{T_0}a_0^3~.
\ee
The constancy of the comoving entropy is essential for the conclusions we shall derive here. We are always considering the semi-classical regime so $\rho_0 \ll (8\pi G_{N})^{-2}$ and $T_0 \ll (8\pi G_{N})^{-1/2}$.

The QES prescription requires the calculation of the entanglement entropy of region $R$. Only few such controlled examples are known without heavy use of symmetries, which the FLRW universe does not possess. We shall therefore use the thermal entropy density of the fluid as a proxy for the entanglement entropy as was also done in \cite{Hartman:2020khs, Bousso:2022gth, Espindola:2022fqb}. At least in high temperatures the entanglement entropy should converge to the thermal entropy \cite{Calabrese:2009qy}. Therefore, our analysis is certainly valid for high enough temperatures, and it remains to be seen how far can one extrapolate it to lower temperatures.

\paragraph{Energy Conditions} General Relativity does not specify which energy momentum tensor should appear in the RHS of Einstein's equations. Therefore, energy conditions are commonly used to avoid certain solutions that may be mathematically correct but not sensible physically. In generality they are specified by the energy momentum tensor and its contraction with certain four-vectors. Since we only deal with perfect fluids, we will use the more simplified version pertaining to a perfect fluid with energy density $\rho$, pressure $p$ and an equation of state $w$ that relates the two $p=w\rho$. The Null Energy Condition (NEC) is $\rho+p=(1+w)\rho \geq 0$, so if $\rho>0$ it means $w\geq-1$. The Dominant Energy Condition (DEC) is $\rho\geq|p|$, i.e. $1\geq |w|$. 

\section{General Island Conditions in FLRW Spacetimes}
\label{section:conditionsRearranged}
Given our choice to work with FLRW spacetimes, we can rearrange the necessary conditions for islands in a simpler way. Recall that we choose to consider spherically symmetrical island candidates centered at $\chi=0$. We denote such an island candidate as $I(\chi)$ and its complement on its Cauchy slice as $G$. The generalized entropy of $I(\chi)$ and $G$ are as follows:
\bea
S_{\mathrm{gen}}(I)&=&S(\eta,\chi)+\frac{A(\eta,\chi)}{4G_{N}}~, \cr
S_{\mathrm{gen}}(G)&=&const.- S(\eta,\chi)+\frac{A(\eta,\chi)}{4G_{N}}~.
\eea
The constant in $S_{\mathrm{gen}}(G)$ is the matter entropy of the entire slice containing $I$ and $G$, but this constant will drop out of the conditions. From these, the four conditions can be written as
\begin{equation} \label{eq:cond_original1}
    \mbox{\textbf{Condition 1:}}~~~~~S(\eta,\chi)>\frac{A(\eta,\chi)}{4G_{N}}~,~~~~~~~~~~~~~~~~~~~~~~~~~~~~~~~~~~~~~
\end{equation}
\begin{equation} \label{eq:cond_original2}
    \mbox{\textbf{Condition 2:}}~~~~~ S'(\eta,\chi)+\frac{A'(\eta,\chi)}{4G_{N}} \geq \mp \left(\dot{S}(\eta,\chi)+\frac{\dot{A}(\eta,\chi)}{4G_{N}}\right)~,~
\end{equation}
\begin{equation} \label{eq:cond_original3}
    \mbox{\textbf{Condition 3:}}~~~~~S'(\eta,\chi)-\frac{A'(\eta,\chi)}{4G_{N}} \geq \mp \left(-\dot{S}(\eta,\chi)+\frac{\dot{A}(\eta,\chi)}{4G_{N}}\right)~,
\end{equation}
\begin{equation} \label{eq:cond_original4}
    \mbox{\textbf{Condition 4:}}~~~~~S^{tot}(\eta)-S(\eta,\chi) \ge \frac{A{(\eta,\chi)}}{4G_{N}}~,~~~~~~~~~~~~~~~~~~~~~~~~~~
\end{equation}
where dot denotes differentiation with respect to conformal time $\eta$, prime denotes differentiation with respect to the radial coordinate $\chi$, and condition 4 applies only to subsets of closed universes.

We can use properties of FLRW spacetimes and the laws of thermodynamics to manipulate the conditions into a more transparent form. In FLRW we have a factorization of the different terms into space and time dependence, $A{(\eta,\chi)}=a^2(\eta)\mathcal{A}(\chi)$, $S(\eta,\chi)=s_{th}(\eta)V(\eta,\chi)\equiv s_c \mathcal{V}(\chi)$. Here $\mathcal{A}$ and $\mathcal{V}$ refer to comoving quantities, while $A$ and $V$ signify physical area and volume. As we have reviewed in section \ref{section:applyingConditions}, the entropy per comoving volume is conserved. Hence, $s_c$ is constant in time and so $S(\eta, \chi)\equiv S(\chi)$. Therefore, the time derivative vanishes $\dot{S}=0$. Finally, it is convenient to rewrite the fourth condition in terms of $S_{tot}$ by adding to it the first condition. Thus the conditions simplify as follows. The first becomes
\begin{equation}
    \label{eq:cond1}
    \mbox{\textbf{Condition 1:}}~~~~~s_c>\frac{a^2(\eta)}{4G_{N}}\frac{\mathcal{A}(\chi)}{\mathcal{V}(\chi)}~.
\end{equation}
Consider conditions (2) and (3). Since $\dot{S}=0$, we can rewrite them as
\begin{equation}
    S'(\eta,\chi) \geq \frac{(\mp\dot{A}\pm A')}{4G_{N}}~,
\end{equation}
 where contrary to common notation, here we mean all possible combinations of signs. This can be rearranged into
 \begin{equation}
     S'(\eta,\chi) \geq \frac{(|\dot{A}|+|A'|)}{4G_{N}}~,
 \end{equation}
and hence conditions 2 and 3 are simply\footnote{$\mathcal{H}\equiv\frac{\dot{a}}{a}$ where dot is a differentiation w.r.t conformal time $\eta$ and $H\equiv\frac{\partial_t a}{a}$ where $t$ is the so called cosmic time.}
 \begin{equation}
    \label{eq:cond23}
    \mbox{\textbf{Conditions 2 and 3:}}~~~~~s_c\geq \frac{a^2(\eta)}{4G_{N}}\left(2|\mathcal{H}|+\frac{|\mathcal{A}'|}{\mathcal{A}}\right)=\frac{a^2(t)}{4G_{N}}\left(2a(t)|H|+\frac{|\mathcal{A}'|}{\mathcal{A}}\right)~.
 \end{equation}
Adding conditions 1 and 4 (for proper subsets of closed universes) yields
 \begin{equation}
     S^{tot}(\eta)=2\pi^2s_c\ge 2\frac{a^2(\eta)\mathcal{A}(\chi)}{4G_{N}}~,
 \end{equation}
 and thus condition 1 + 4 is
 \begin{equation}
    \label{eq:cond1p4int}
    \mbox{\textbf{Condition 1 + 4:}}~~~~~s_c\geq\frac{a^2(\eta)}{4G_{N}}\frac{\mathcal{A}(\chi)}{\pi^2}~.
 \end{equation}

Before going into specific examples we can see what conditions are more likely to be satisfied and where we may find potential obstacles. We have managed to phrase all conditions in terms of a inequalities on the comoving entropy density, $s_c>a^2/4G_{N} f$ where $f$ is the fudge factor that is different between the different conditions. From this general expression, we can understand why time-symmetric slices are good candidates for islands. This is because the existence of such a slice means that $a$ is bounded, and given a concentration of enough entropy, the conditions will be fulfilled. Moreover, given that the difference between the conditions is the fudge factor $f$, we expect that in various cases a single condition to encompass all others.\footnote{One would usually call this condition a sufficient condition, as its fulfillment implies that all others are fulfilled as well. However, this is not a sufficient condition for the existence of islands, that requires further checks (see below). It is only a sufficient condition for a viable island candidate. To avoid confusion, we use the name encompassing condition, meaning that it captures all other necessary conditions.}

\section{Islands on Time-Symmetric Slices} 
\label{section:TSS}
In this section, we begin to investigate where the necessary conditions can be satisfied in a more general cosmology. First, we do not restrict ourselves to models with perfect fluids, allowing more general contents in an FLRW universe. Because this complicates the question, we start by restricting to time-symmetric slices, \textit{i.e.} slices in which the scale factor has a minimum or maximum. The QES prescription relies on extremization of surfaces. Time-symmetric slices are an extremum in the time coordinate. Therefore, these slices are natural places to look for islands. Indeed, with the exception of an entire closed universe, the viable island candidates found in Refs. \cite{Hartman:2020khs, Espindola:2022fqb, Bousso:2022gth} were on or near time-symmetric slices.

Consider a time-symmetric slice and its vicinity, where $\mathcal{H}=0$ or extremely small. Hence,
\bea
\mbox{\textbf{Condition 1:}}~~~~~s_c\geq\frac{a_{ts}^2}{4G_{N}}\frac{\mathcal{A}(\chi)}{\mathcal{V}(\chi)}~,~~~~~~~~~\cr
\mbox{\textbf{Conditions 2 and 3:}}~~~~~s_c\geq \frac{a_{ts}^2}{4G_{N}}\frac{|\mathcal{A}'|}{\mathcal{A}}~,
\eea
where $a_{ts}$ denotes the scale factor at the time-symmetric slice.
For flat universes, $\chi$ can be arbitrarily large, and $\mathcal{A}/\mathcal{V}$ and $\mathcal{A}'/\mathcal{A}$ can thus be arbitrarily small by taking large enough $\chi$. Therefore, for flat universes, an arbitrary small amount of entropy $s_c$ will still fulfill all conditions for large enough $\chi$. Hence, time-symmetric slices in flat universes always contain viable island candidates. This is regardless of any other detail such as number of fluids, energy component etc. It is based solely on the constancy of $s_c$, and the existence of a time-symmetric slice. A specific example of such a universe was found in \cite{Hartman:2020khs}, where a flat universe with radiation and a negative CC was considered.

For open universes $\mathcal{A}/\mathcal{V}$ and $\mathcal{A}'/\mathcal{A}$ are bounded from below and asymptote to 2 for large $\chi$, resulting in the encompassing condition:
\be \label{eq:time-openencomp}
s_c\geq \frac{a_{ts}^2}{2G_{N}}~.
\ee
For closed universes, the whole manifold is always an island, \cite{Bousso:2022gth}. Contrary to the flat and open case, the quantities $\mathcal{A}/\mathcal{V}$ and $\mathcal{A}'/\mathcal{A}$ are not monotonic in $\chi$, so there is no general $\chi$ beyond which all conditions reduce to a single encompassing one.  Beyond that we can consider a specific region of certain $\chi$ where the conditions are more simple to satisfy. Such a region is  $\chi \simeq \pi/2$ on a time-symmetric slice, since \eqref{eq:cond23} is trivially fulfilled. The rest of the conditions reduce to the encompassing condition 
\be \label{eq:time-closedencomp}
s_c\geq \frac{a_{ts}^2}{\pi G_{N}}~.
\ee
 From this we can see a reason why closed universes are in some sense more island-friendly. In an ever expanding universe, $a$ grows without bound and therefore will always violate this condition after a long enough time. Ever expanding closed universes will require $w<-1/3$ and/or a positive CC, which are not trivial. In contrast, the violation of the inequality may not happen in a universe with a bounded $a_{ts}$, meaning it is ``easier" to fulfill this condition. 

The above analysis depends on the existence of a time-symmetric slice. The matter and energy content of the universe affects whether such a slice can exist. Exotic matter or energy components can be introduced to produce such a slice, e.g. \cite{Artymowski:2019cdg}. It is, however, standard to treat the matter contents of an FLRW universe as a collection of perfect fluids, and possibly a cosmological constant and spatial curvature. Let us call such a universe, one that contains any number of perfect fluids, ``conventional." We assume that all energy components are perfect fluids and further that all energy densities of all fluids are positive definite. We still restrict our search to time-symmetric slices, the existence of which are determined by the first Friedmann equation:
\be
0=\frac{8\pi G_{N}}{3}\sum_i \rho_{0,i} \left(\frac{a_{ts}}{a_0}\right)^{-3(1+w_i)} \pm \frac{1}{t_{\Lambda}^2}-\frac{k}{a_{ts}^2}~,
\ee
where $\rho_{0,i}\geq0$ is the energy density of various fluids at the turnaround time and $\pm$ is the sign of the CC. Hence, for flat and open conventional universes, a solution to the above equation exists only if there is a negative CC. For a closed universe a time-symmetric slice can exist with any CC depending on the value of $a_{ts}$. We tabulate the islands in general and specifically in the case of conventional time-symmetric slices in Table \ref{table1}. The islands found in \cite{Hartman:2020khs, Bousso:2022gth, Espindola:2022fqb} correspond to the conventional time-symmetric slices.

\begin{center}
\begin{table}[H] \label{table1}
\caption{Summary of viable island candidates on time-symmetric slices. The second column corresponds to island candidates in the vicinity of time-symmetric slices without limiting ourselves to perfect fluids with positive definite energy densities, while the third column corresponds to the inclusion of this limitation. The flat and open universe cases here include all possible spatial time-symmetric islands $\chi \gg 1$, while the closed universe is limited to the $\chi\simeq\pi/2$ case.} 
\hspace{1.5 cm}
\begin{tabular}{|l|l|l|}
\hline
\bf{Case} & \bf{General Time sym.} & \bf{Conventional Time sym.}  \\ \hline
$k=0$ & Always & If $\Lambda<0$ exists \\
\hline
$k=-1$ & If $s_c>\frac{a_{ts}^2}{2G_{N}}$ & If $s_c>\frac{a_{ts}^2}{2G_{N}}$ and $\Lambda<0$ exists \\
\hline
$k=1$ & If $s_c>\frac{a_{ts}^2}{\pi G_{N}}, \quad \chi \simeq \pi/2$ & Any $\Lambda$, if $s_c>\frac{a_{ts}^2}{\pi G_{N}}, \quad \chi \simeq \pi/2$ \\
 \hline
\end{tabular}
\end{table}
\end{center}

\section{Beyond Time-Symmetric Slices: Flat Universes}
\label{section:flat}
Let us now deviate from time-symmetric slices and their vicinity. We wish to investigate where the necessary conditions for islands are satisfied in flat universes with a general perfect fluid. Consider a flat FLRW universe filled with a perfect fluid with an equation of state $w$. We consider spherical islands candidates that exist far away from singularities, so as to stay in the semiclassical regime. Hence, we are interested in cases where $\eta \gg 1$ and $\rho \ll (8\pi G_{N})^{-2}$. The comoving volume and area are the standard Euclidean expressions for a sphere, $\mathcal{V}(\chi)=\frac{4\pi}{3}\chi^3$ and $\mathcal{A}(\chi)=4\pi \chi^2$. 
The scale factor and Ricci scalar may be written as, respectively,
\begin{equation}
\label{eq:aflat}
a=a_0(t/t_0)^{2/(3+3w)}=a_0(\eta/\eta_0)^{2/(1+3w)}~,
\end{equation}
\begin{equation}
\mathcal{R}=\frac{6}{a^2}\left(\dot{\mathcal{H}}+\mathcal{H}^2\right)=6\left(\frac{dH}{dt}+2H^2\right)\sim t^{-2} \sim \eta^{-6(1+w)/(1+3w)}~.
\end{equation}
Hence, at $t=0$ we will hit the Big Bang singularity. However, this is not enough. To trust our semi-classical analysis we need to be far away from Planck energy densities, which amounts to
\be
\rho_0\ll (8\pi G_{N})^{-2} \Leftrightarrow t_0 \gg \sqrt{8\pi G_{N}}~.
\ee
Here $t_0$ is an arbitrary time of normalization and we specify our results with respect to it. Finally, in a flat universe with a single fluid $\rho_0$ and $t_0$ are related via the Friedmann equation at $t_0$
\be \label{eq:Friedmannt0}
\left(\frac{2}{3(1+w)t_0}\right)^2=\frac{8\pi G_{N}}{3}\rho_0~.
\ee

\subsection{Island candidates at $t=t_0$}
Let us first consider the possibility of island candidates at $t=t_0$, where $t_0$ is an arbitrary normalization time. This possibility answers the following question: Assuming we are allowed to tune the temperature $T_0$ and the energy density $\rho_0$ at a given time $t_0$ in the FLRW universe, and that the universe maintains thermal equilibrium, can an island exist?
Substituting \eqref{eq:aflat} into conditions \eqref{eq:cond1}, \eqref{eq:cond23} and $t=t_0$ yields

\begin{gather} \label{eq:cond1_at_t0}
\mbox{\textbf{Condition 1:}}~~~~~s_c>\frac{3a_0^2}{4G_{N}\chi}~,~~~~~~~~~~~~~~~~~~~~~~~~~~~~~~~~~~~~~~\\
\label{eq:cond23_at_t0}
\mbox{\textbf{Conditions 2 and 3:}}~~~~~s_c\geq \frac{a_0^3}{4G_{N}} \left[\frac{4}{3(1+w)|t_0|}+\frac{2}{a_0\chi}\right]~.
\end{gather}
Using \eqref{eq:Friedmannt0} and the expression for the comoving entropy density \eqref{eq:sc} yields the following inequalities:

\begin{gather}
\mbox{\textbf{Condition 1:}}~~~~~\frac{(1+w)G_{N}\rho_0a_0}{T_0}>\frac{3}{4\chi}~,~~~~~~~~~~~~~~~~~~~~~~~~~~~~~~~\\
\mbox{\textbf{Conditions 2 and 3:}}~~~~~\frac{(1+w)G_{N}\rho_0a_0}{T_0}>\frac{1}{2\chi}+a_0\sqrt{\frac{2\pi G_{N} \rho_0}{3}}~.
\end{gather}

We have the following dimensionful parameters $G_{N},\rho_0,T_0$ and $\chi_{0,phys.}\equiv a_0\chi$, the spatial size of the island candidate, and some $\mathcal{O}(1)$ numbers which are not important at the moment. There are two possible regimes for island candidates: $\chi_{0,phys.}^{-1} \gg \sqrt{G_{N}\rho_0}$ and $\sqrt{G_{N}\rho_0} \gg \chi_{0,phys.}^{-1}$. If $\chi_{0,phys.}^{-1} \gg \sqrt{G_{N} \rho_0}$, then the first condition simplifies to
\begin{equation}
    \label{eq:chiphys}
    G_{N}\rho_0 > \mathcal{O}(1)\, \frac{T_0}{\chi_{0,phys.}}~,
\end{equation}
and the second term in the ``2 and 3" condition is negligible, meaning it is encompassed by the first. Thus satisfaction of \eqref{eq:chiphys} will ensure a viable island candidate. The other regime is if $\sqrt{G_{N}\rho_0} \gg \chi_{0,phys.}^{-1}$, which implies that \eqref{eq:chiphys} is trivial and the ``2 and 3" condition simplifies to
\begin{equation}
    \label{eq:T0square}
    \frac{G_{N}\rho_0}{T_0}>\mathcal{O}(1)\sqrt{G_{N} \rho_0} \quad \Rightarrow  \quad G_{N}\rho_0 > \mathcal{O}(1) \, T_0^2~.
\end{equation}
Then \eqref{eq:T0square} will ensure a viable island candidate. 

Given freedom to tune $\rho_0$ and $T_0$ as independent energy scales, we see that viable island candidates will appear for practically any fluid and will be of any size as long as $T_0$ and $\rho_0$ obey
\be
G_{N} \rho_0>\mathcal{O}(1) \max \left\{T_0^2, \frac{T_0}{\chi_{0,phys.}}\right\}~.
\ee
The statement also has a nice interpretation in terms of the Hubble parameter since $H\sim \sqrt{G_{N}\rho}$:
\be
H_0>\mathcal{O}(1) \max \left\{T_0, \sqrt{\frac{T_0}{\chi_{0,phys.}}}\right\},
\ee
where $H_0$ is the Hubble parameter at the time of normalization $t_0$, and not today, as is usually denoted.
\begin{figure*}
    \centering
    \includegraphics[width=.8\textwidth]{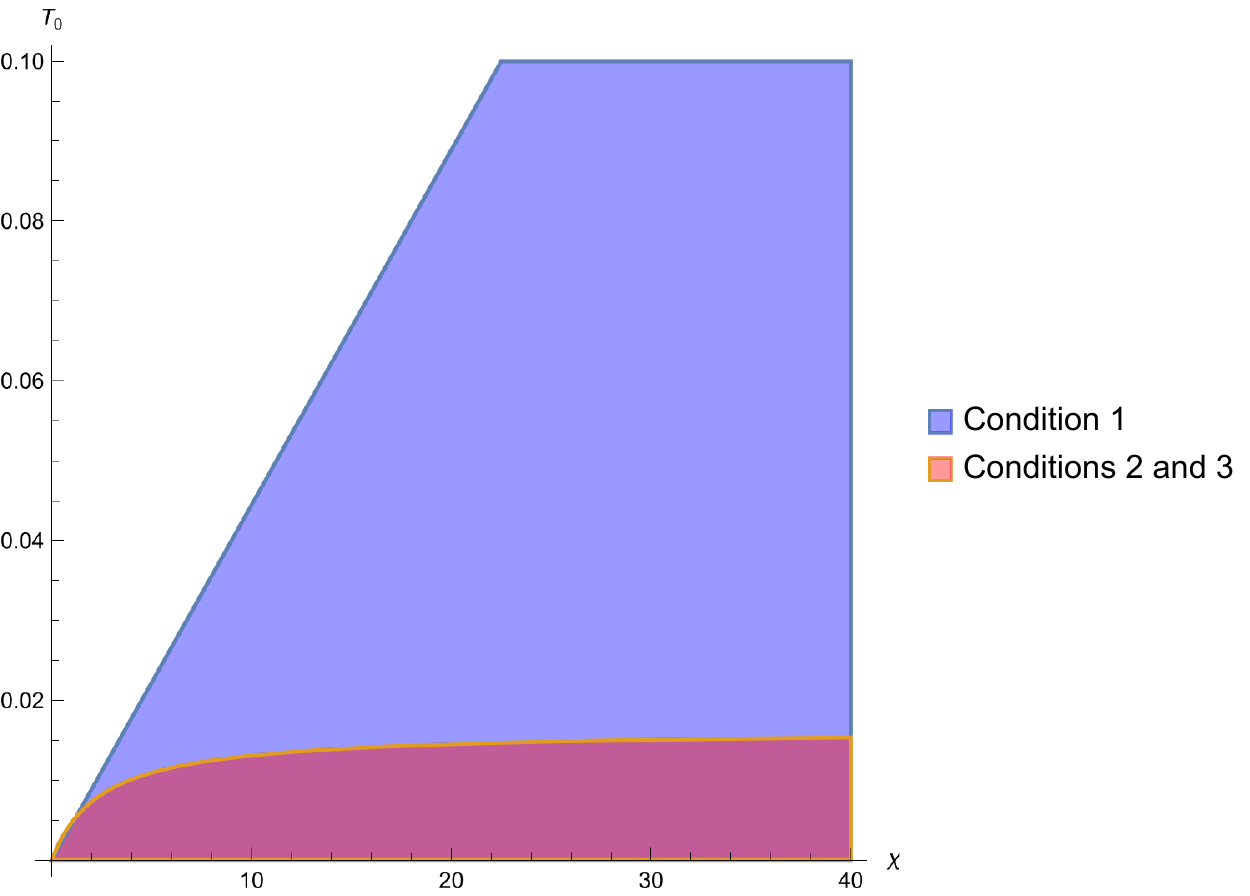}
    \caption{A plot of $T_0$ vs $\chi$ for $w=2/3$, $\rho_0=2\times10^{-4}$ in natural units with the normalization $a_0$ set to 10. The conditions at a given $\chi$ are only satisfied for a small enough $T_0$, but for any $\chi$ there is always such a $T_0$ for which the conditions are all satisfied.}
    \label{fig:TempChi}
\end{figure*}

As an example of the relationship between $T_0$ and $\chi$ for a chosen $\rho_0$, we show in Figure \ref{fig:TempChi} a plot of $T_0$ vs $\chi$ for $w=2/3$, $\rho_0=2\times10^{-4}$. At any $\chi$, there is a range of small enough $T_0$'s such that the conditions are simultaneously satisfied.

Radiation is a special case where $\rho_0=c_{th} T_0^4$, where $c_{th}$ is roughly the number of degrees of freedom. Suppose we take $\chi_{phys}$ arbitrarily large to make \eqref{eq:chiphys} trivial. Substituting $\rho_0=c_{th} T_0^4$ into \eqref{eq:T0square} 
shows that the inequality is not fulfilled unless $T_0$ is Planckian or $c_{th} \gg 1$. In other words, for radiation $\rho_0$ and $T_0$ are not tuned appropriately to allow for viable island candidates.\footnote{Another example is degenerate fermions, which have $\mu \gg T$,
$\rho_0=\frac{c_{th}}{8\pi^2} \mu^4$, $n_0=\frac{c_{th}}{6\pi^2} \mu^3$, $p_0=\frac{c_{th}}{24\pi^2} \mu^4$ \cite{Kolb:1990vq}.
We therefore have to reintroduce $\mu$ into the expression of $s_c$. Calculating $s_c$ from \eqref{eq:sth} shows that $s_c$ vanishes, which means the conditions are not satisfied.}

\subsection{Islands away from $t=t_0$}
We showed in the previous section that if $\rho_0$ and $T_0$ are separate free parameters, we can always tune them to create a viable island candidate on the normalization slice $t_0$. In the real world, however, we do not have this freedom. It is possible that $\rho_0$ and $T_0$ in our universe do not fulfill the conditions at the present time. Thus in this section we investigate whether a universe without islands at $t_0$ can evolve into or have evolved from one with islands.

The simpler case is to treat the normalization as the initial time, and then look for islands after this initial time $t \gg t_0$ because semiclassical regime is guaranteed: $t \gg t_0 \gg \sqrt{8\pi G_{N}},\quad \rho<\rho_0\ll (8\pi G_{N})^{-2}$. Assume at the normalization time $t_0$, the universe with temperature $T_0$ and energy density $\rho_0$ did not have a viable island candidate. By considering conditions \eqref{eq:cond1}, \eqref{eq:cond23} and \eqref{eq:sc} with their time dependence, we get the following conditions:
\begin{gather}
\mbox{\textbf{Condition 1:}}~~~~~G_{N} \rho_0>\frac{3}{4(1+w)}\frac{T_0}{\chi_{phys.}}\left(\frac{a}{a_0}\right)^{3} \label{eq:Grho1}~,~~~~~~~~~~~~~~~~~~~~~~~\\
\mbox{\textbf{Conditions 2 and 3:}}~~~~~G_{N}\rho_0>\frac{T_0}{2(1+w)}\, \left(\frac{a}{a_0}\right)^{3}\left[|H|+\frac{1}{\chi_{phys.}}\right] \label{eq:Grho23}~,
\end{gather}
where $\chi_{phys.}\equiv a \chi$ is the physical radius of the island candidate. 
These expressions are valid for both contracting and expanding universes.
Since $0\leq \chi<\infty$, for any given time $t$, for any small amount of energy density of $G_{N}\rho_0$ (or equivalently comoving entropy $s_c$) there always exists $\chi_{phys.}$ such that \eqref{eq:Grho1} is fulfilled, and the $\chi$ dependent term in 
\eqref{eq:Grho23} is negligible.\footnote{Here we do not consider the case where the $\chi$ term is the dominant one because an island candidate in such a case will mean that the it already existed at $t=t_0$, contrary to our initial assumption in this section. It reduces back to the question of an island at $t=t_0$, which we have already analyzed.} Therefore the $H$-dependent part in \eqref{eq:Grho23} is our encompassing condition. Using the Friedmann equation \eqref{eq:Friedmann1} and \eqref{eq:rhoi}, we can rewrite \eqref{eq:Grho23} as
\begin{gather}
\label{eq:2wflat}
\mbox{\textbf{Conditions 2 and 3:}}~~~~~G_{N}\rho_0>\frac{2\pi}{3(1+w)^2}\, T_0^2\left(\frac{a}{a_0}\right)^{3(1-w)}~.
\end{gather}
To ensure islands with $a \gg a_0$, we therefore need\footnote{
Since in many bouncing models such as the ekpyrotic scenario we have $w\gg1$, we get that islands in such models are ubiquitous at early stages of the contraction.} 
\be
\boxed{w\geq 1}~.
\ee

Thus for positive energy density $\rho>0$, the necessary conditions for islands imply a violation (or saturation) of the Dominant Energy Condition (DEC), which stipulates $|w|\leq 1$. In certain cases such as radiation, one can further derive the exact relation between $s_c,\rho_0,T_0$ and the number of degrees of freedom, $c_{th}$ e.g. \cite{Hartman:2020khs}, but that is unnecessary. The point is that for a finite $s_c$ there will always be a viable island candidate for $t \gg t_0$ and a large enough $\chi$ if $w>1$. Substituting the exact value of $s_c$ from \eqref{eq:sc} and \eqref{eq:aflat}, one gets the exact time when the island may form:
\be
t>t_0\left(\frac{1}{2\pi T_0t_0}\right)^{(1+w)/(1-w)}~.
\ee

For the special case of $w=1$, the issue of a viable island candidate becomes a quantitative question regarding the exact values of $s_c$, $\rho_0$, etc. 
Considering again a large enough $\chi$ such that \eqref{eq:Grho1} is fulfilled and substituting \eqref{eq:sc} into \eqref{eq:2wflat} for $w=1$, we get a condition on the temperature $T_0$:
\be
T_0<\sqrt{\frac{24 G_{N} \rho_0}{\pi}}~,
\ee
for a viable island candidate. 

Let us now consider the more delicate case, where our normalization time is $t_0$, and we would like to investigate the possibility of islands to its past while staying in the semiclassical regime, i.e. $t_0 \gg t\gg \sqrt{8\pi G_{N}}$. Condition \eqref{eq:cond1} can still be fulfilled for any time by taking large enough $\chi$. We then need to consider \eqref{eq:cond23}. Our assumption is that at $t_0$ there is no island, and we wish to check whether at $t\ll t_0$ there may be one. We get the converse condition, that the DEC has to be fulfilled:
\bea
\frac{a_0^3}{3(1+w)G_{N}}\frac{|t_0|^{(1-w)/(1+w)}}{t_0^{2/(1+w)}}\geq  s_c\geq \frac{a_0^3}{3(1+w)G_{N}}\frac{|t|^{(1-w)/(1+w)}}{t_0^{2/(1+w)}}~,\cr
\Rightarrow |w|<1, \quad \frac{1}{2\pi t_0}\left(\frac{t_0}{t}\right)^{(1-w)/(1+w)}\geq T_0\geq  \frac{1}{2\pi t_0}~. \label{eq:backwardevolution}
\eea
The upper bound on $s_c$ comes from requiring that there is no island at $t=t_0$, and the lower bound from requiring that there is an island at $t\ll t_0$.

On top of that we have to show that we are still in the semiclassical regime. If there is an additional energy scale, say $\rho_0 \neq T_0^4$, then it is simple to fulfill the inequalities by choosing the appropriate energy scale. 
An immediate example is (non-relativistic) dust with an equation of state $w=0$ and particle mass $m$. Equation \eqref{eq:backwardevolution} gives
\be
\frac{1}{2\pi t} \geq T_0 \geq \frac{1}{2\pi t_0}~.
\ee

\begin{figure*}
    \centering
    \includegraphics[width=.8\textwidth]{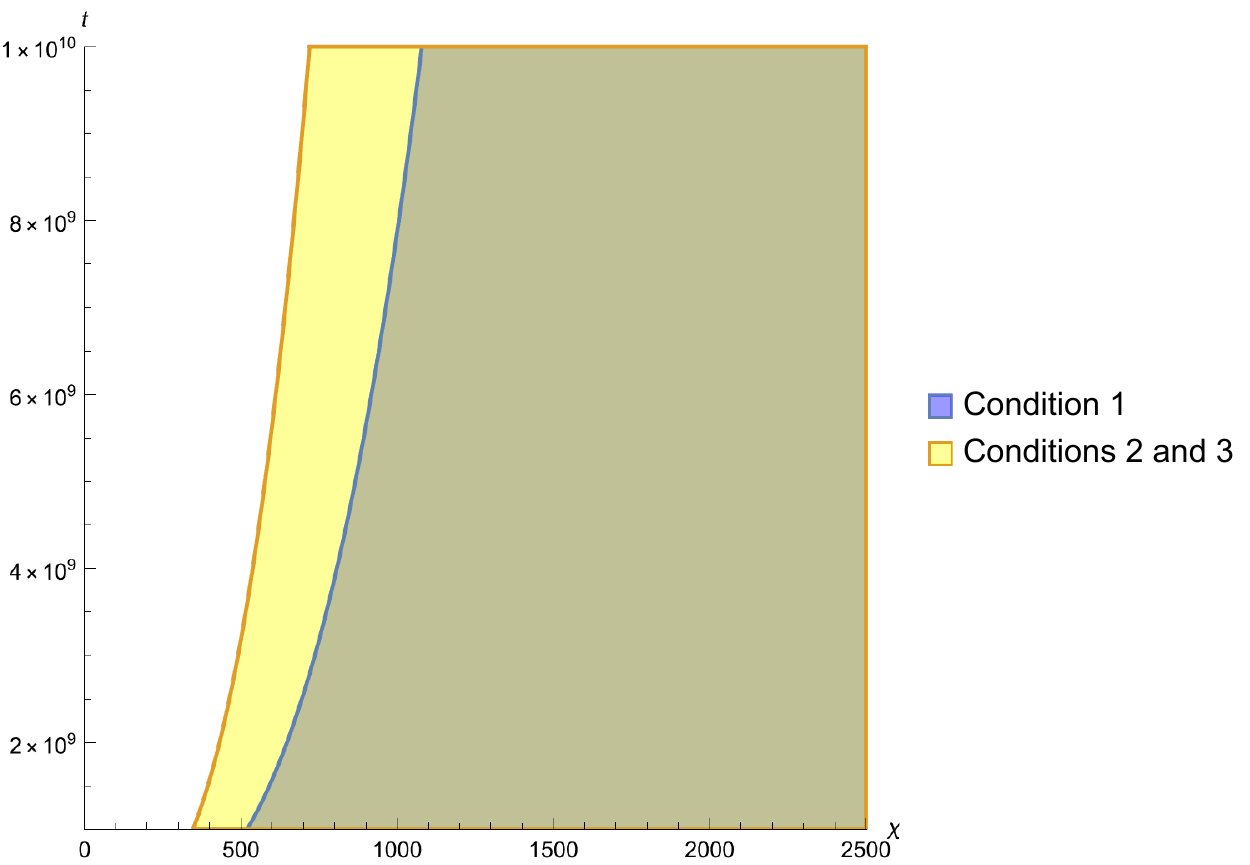}
    \caption{Regions satisfying the island conditions are shown for a flat universe with $w=3.2,~\rho_0=10^{-6},~T_0=10^{-4}$.}
    \label{fig:plotGerenal1}
\end{figure*}

\begin{figure*}
    \centering
    \includegraphics[width=.8\textwidth]{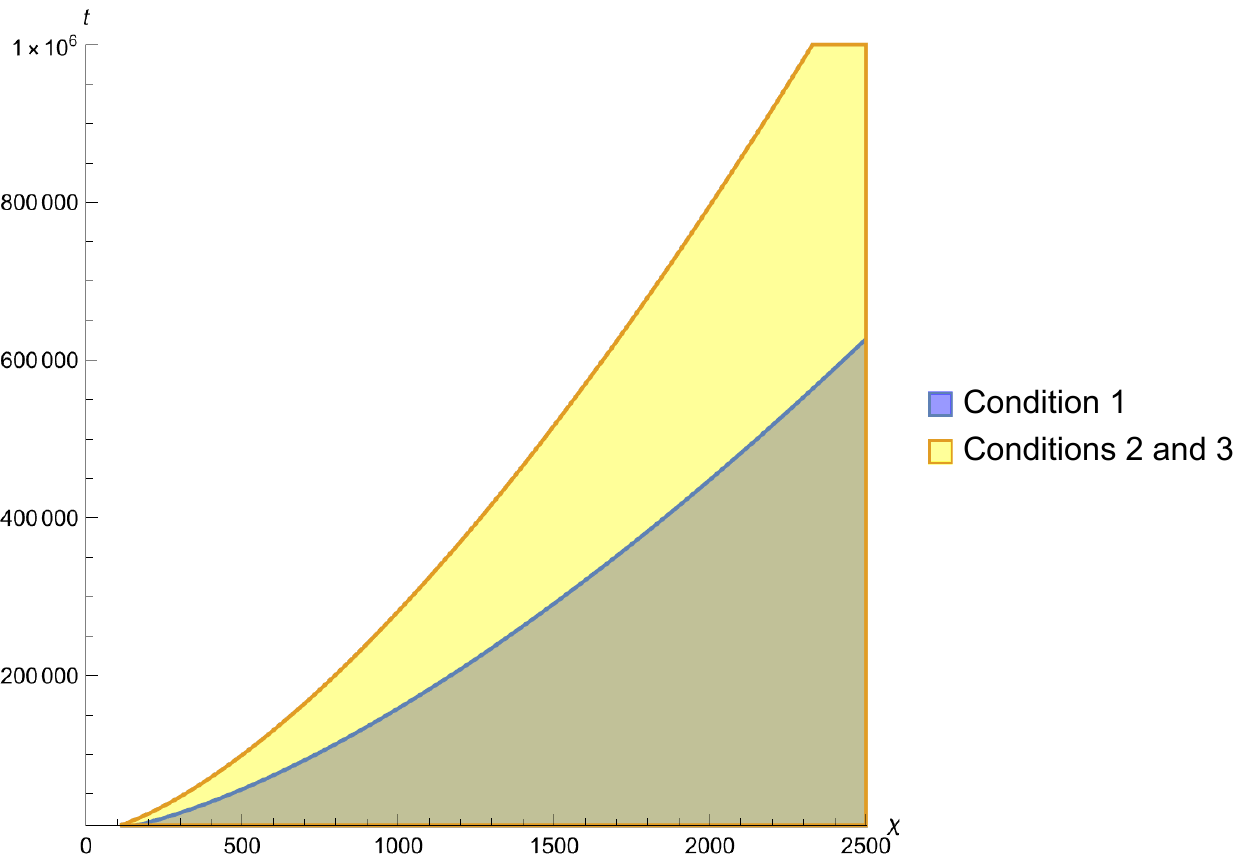}
    \caption{Regions satisfying the island conditions are shown for a flat universe with $w=1,~\rho_0=10^{-6},~T_0=10^{-4}$.}
    \label{fig:plotGerenal2}
\end{figure*}

Having discussed the conditions at $t_0$ and before and after $t_0$, we now plot two examples of the regions satisfied by the conditions: one with $w=3.2,~ \rho_0=10^{-6},~T_0=10^{-4}$ (Figure \ref{fig:plotGerenal1}) and the other with $w=1,~\rho_0=10^{-6},~T_0=10^{-4}$ (Figure \ref{fig:plotGerenal2}) in natural units $G_{N}=1$. In these plots we substituted the exact expressions for \eqref{eq:cond1}, \eqref{eq:cond23}. They show that for the parameters chosen, a given time slice always has some minimum $\chi$ for which all the necessary conditions are fulfilled.

The conditions here are only necessary conditions for islands; satisfying all conditions does not guarantee that a particular region is an island. To check that a viable island candidate is in fact an island, we must look at the reference region of interest and confirm that application of the QES prescription forces the island candidate into that region's entanglement wedge. This is straightforward in the model discussed in section \ref{section:preliminaries}, where $M$ and $M_R$ are entangled in a thermofield-double-like state. In this model, a viable island candidate $I(\chi)$ for the flat space scenarios we have so far discussed will be an island if we select the matching region $R$ in the nongravitational spacetime $M_R$. This is because (i) the candidate satisfies the homology constraint of the QES prescription, (ii) satisfying the original conditions 2 and 3 implies the candidate is quantum extremal, and (iii) including the island in the entanglement wedge of $R$ is generalized-entropy-minimizing. Regarding (iii), because of the entanglement structure in the thermofield-double-like model, decreasing the size of the island would increase the matter entropy in the entanglement wedge by an amount that goes with the volume (from the entangled pairs in $R$ that are no longer purified by the island) while decreasing the area contribution by an amount that, naturally, goes with the area. Increasing the size of the island would increase both the matter and area contributions. Thus the island candidate must be included in the entanglement wedge of $R$, and therefore $I(\chi)$ is an island. This argument holds for all the flat space scenarios we have discussed thus far.

\section{Beyond Time-Symmetric Slices: Closed and Open Universes}
\label{section:closedandopen}
In this section we rearrange the conditions in the new forms we derived in section \ref{section:conditionsRearranged} for universes with nonzero curvature and a single general fluid. A drawback regarding geometries with spatial curvature is that generically we do not have an explicit time dependence and we can only give results in terms of the scale factor $a$.

First consider closed universes ($k=+1$) with a general perfect fluid. The comoving volume of a sphere in a closed geometry is $\mathcal{V}(\chi) = \pi(2\chi-\sin2\chi)$, so the matter entropy of an island candidate of comoving radius $\chi$ is
\begin{equation}
    S=s_c \pi (2\chi-\sin2\chi)~.
\end{equation}
The corresponding comoving area and its spatial derivative are $\mathcal{A}=4\pi \sin^2\chi$ and $\mathcal{A}'=4\pi \sin 2\chi$ respectively. We substitute the closed geometry into the simplified conditions \eqref{eq:cond1}, \eqref{eq:cond23}, \eqref{eq:cond1p4int} to obtain
\begin{equation}
    \mbox{\textbf{Condition 1:}}~~~~~s_c\geq \frac{a^{2}}{G_{N}}\frac{\sin^2\chi}{2\chi-\sin 2\chi}~,~~~~~~~~~~~~~~~~~~
\end{equation}
\begin{equation}
    \mbox{\textbf{Conditions 2 and 3:}}~~~~~s_c \geq \frac{a^2(\eta)}{2G_{N}}\left(|\mathcal{H}|+|\cot \chi| \right)~,
\end{equation}
\begin{equation}
    \mbox{\textbf{Condition 1 + 4:}}~~~~~s_c\geq\frac{a^2}{G_{N}}\frac{\sin^2{\chi}}{\pi}~,~~~~~~~~~~~~~~~~~~~
\end{equation}
where Conditions 2 and 3 together come from Eq.~\eqref{eq:cond23}. These are the general conditions for island candidates in closed universes. There are two potential obstacles for islands. First, if $|\cot \chi|$ diverges. Second, if the scale factor grows without bound. 
We can overcome the first difficulty by looking for islands around $\chi\simeq \pi/2$. In such a case, the conditions simplify to the encompassing condition
\be \label{eq:cond_closed_enc}
s_c\geq\frac{a^2}{\pi G_{N}}~,
\ee
provided $\mathcal{H}< 2/\pi$, which is generally the case in an expanding universe. 

Considering a closed universe with a perfect fluid and a CC, there are three possible regimes: perfect fluid domination, curvature domination and CC domination. For the latter two, we do not expect to find anything different than the analysis carried out in \cite{Bousso:2022gth}, except that the parameters of the fluid are different from radiation. In the case of fluid domination,  $G_{N}\rho_{0}/a^{3(1+w)} \gg \{t_{\Lambda}^2, a^{-2}\}$, the analysis resembles that of the previous section. 
In an expanding universe where $a$ is monotonically growing, the only way to satisfy \eqref{eq:cond_closed_enc} is if it is satisfied at the time of normalization where $a=a_0$, when we evolve backwards and $a<a_0$, or for some finite time in the future where $a>a_0$. Once again we get a condition relating the energy density $\rho_0$ and the temperature $T_0$. Considering island candidates at the time of normalization where $a=a_0$, \eqref{eq:cond_closed_enc} reduces to
\be
G_{N} \rho_0>\frac{T_0}{(1+w)\pi a_0}~,
\ee
while for backward time evolution we have
\bea
\frac{a^2}{\pi G_{N}}\leq s_c \leq \frac{a_0^2}{\pi G_{N}}\cr
\Rightarrow \frac{T_0}{(1+w)\pi a_0}\left(\frac{a}{a_0}\right)^2\leq G_{N}\rho_0 \leq \frac{T_0}{(1+w)\pi a_0}~.
\eea
Notice that we did not need to use the value of the CC. Hence, islands which are subset of the closed universe manifold can exist with any type of CC, provided that the universe is dominated by the perfect fluid for long enough time.

For $\chi\neq \pi/2$, there is no simplified treatment when we move away from time-symmetric slices. The inequalities in general then become:
\be
s_c\geq \frac{a^2}{G_{N}}\times \max\left\{\frac{\sin^2\chi}{2\chi-\sin 2\chi},\,\frac{|\mathcal{H}|+|\cot \chi|}{2},\frac{\sin^2 \chi}{\pi}\right\}~.
\ee

There is a particular closed universe model that warrants attention as a specific example. Notice that until now all known examples of islands contained spacetime singularities. It is interesting to consider islands in space time without singularities at least classically. We  therefore apply our findings to the ``Simple Harmonic Universe" scenario, \cite{Graham:2011nb}. This scenario consists of a perfect fluid $-1<w\leq -1/3$, positive spatial curvature $k=+1$ and a negative CC. It does not have a singularity at the background level and is classically stable for certain range of parameters\footnote{To be precise, it is classically stable for $a_{max}/a_{min} \sim \mathcal{O}(1)$, and for $a_{max}/a_{min} \gg 1$ it is stable for many cycles until the approximation scheme in \cite{Graham:2011nb} breaks down.}. Specifically for $w=-2/3$ there is an analytic solution for the scale factor. The scale factor is periodic, taking the form
\begin{equation}
    a(t)=\frac{\rho_0}{2|\Lambda|}+a_0\cos(\omega t + \psi)~,
\end{equation}
where $\omega\equiv\sqrt{\frac{8\pi G_{N}|\Lambda|}{3}}$ and $a_0\equiv\frac{1}{2|\Lambda|}\sqrt{\frac{-3|\Lambda|}{2\pi G_{N}}+\rho_0^2}$.

From the thermodynamic relation, Eq.~\eqref{eq:sthw}, with $w=-2/3$, the comoving entropy density obeys
\begin{equation}
    s_c\sim \rho_0^3~.
\end{equation}

\begin{figure*}
    \centering
    \includegraphics[width=.8\textwidth]{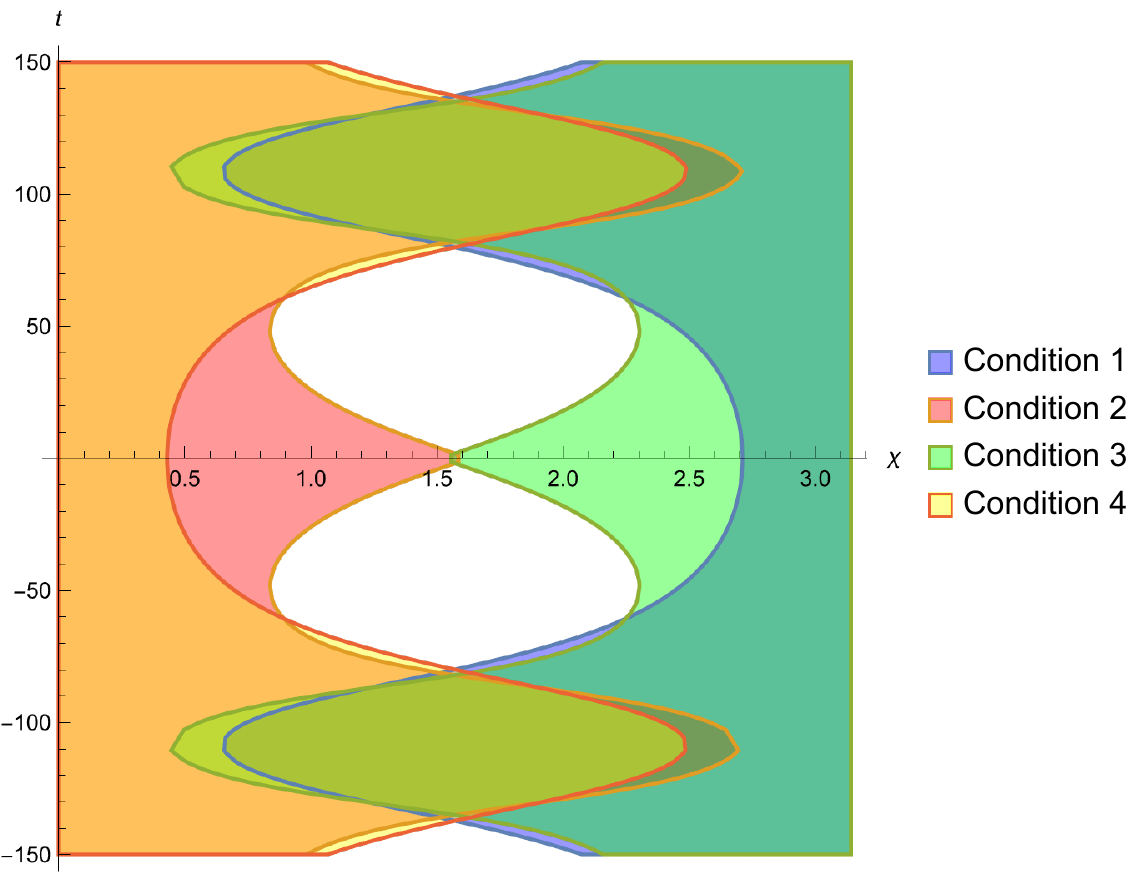}
    \caption{Regions satisfying the four island conditions are shown for the simple harmonic universe, with $\rho_0=0.01, |\Lambda|=0.0001$, $T_0=1.6\times10^{-5}$. We choose to phrase the conditions here according to their original form in \eqref{eq:cond_original1}, \eqref{eq:cond_original2}, \eqref{eq:cond_original3}, \eqref{eq:cond_original4} due to their nice symmetrical properties.}
    \label{fig:plotSHU}
\end{figure*}

The simple harmonic universe has two classes of time-symmetric slices: when $a(t)$ reaches a minimum or a maximum. The scale factor at these times can be found by setting $\dot{a}$ to $0$ in the Friedmann equation, which yields:
\begin{equation}
    a_{ts}=\frac{\rho_0}{2|\Lambda|}\left(1\pm\sqrt{1-\frac{3|\Lambda|}{2\pi G_{N}\rho_0^2}}\right)~,
\end{equation}
where the $+$ or $-$ corresponds to $a_{max}$ or $a_{min}$ respectively. The semi-classical regime requires $a(t)\gg l_P$, which in this case corresponds to $a_{min}\gg l_P$. For $a_{min}\gg1$ (in natural units) we need to pick a CC and $\rho_0$ such that at all times the energy density is smaller than Planckian. This can be achieved for instance with $\rho_0=0.01 M_{P}^3$ and $|\Lambda|=10^{-4}M_{P}^4$, that results in $a_{min}\simeq 13/M_{P}$ and a $\rho_0$ that is always much smaller than Planckian. The complete spacetime for this parameter regime is plotted in Figure \ref{fig:plotSHU}. Here $\chi=0$ corresponds to a time at which $a(t)$ is at a maximum. There are regions of 4-way overlap around the times at which $a(t)$ reaches a minimum. Furthermore, entire Cauchy slices of this universe are always viable island candidates, which follows directly from the results of \cite{Bousso:2022gth}.

As before, we can check in the thermofield-double model whether these viable island candidates are in fact islands for some reference region $R$. The argument is the same, but now subsets of this universe compete the with the entire Cauchy slice, which is always a viable island candidate. A viable island candidate that is a subset of the Cauchy slice $\Sigma_M$ will be an island for a region $R$ of equal size and location on $\Sigma_R$. The entire Cauchy slice will be an island if $R$ is more than half of $\Sigma_R$ and condition 4 is violated.
Hence, in our model, the time-symmetric slices at $a_{min}$ in the Simple Harmonic Universe scenario are actually islands.

Next consider open universes ($k=-1$). For such universes, the geometric factors are $\mathcal{V}=\pi(\sinh 2\chi-2\chi),\, \mathcal{A}=4\pi \sinh^2 \chi$, and $\mathcal{A}'=4\pi \sinh 2\chi$. One can substitute these expressions and look for potential islands. The best chance is in the large $\chi$ limit, as the geometric factors asymptote to a constant value $\mathcal{A}'/\mathcal{A}\rightarrow 2$ and $\mathcal{A}/\mathcal{V}\rightarrow 2$. In the semiclassical regime $|\mathcal{H}| \ll 1$ in natural units. Hence, all conditions collapse to a single encompassing condition
\be \label{eq:cond_open_enc}
s_c\geq \frac{a^2}{2G_{N}}~.
\ee

Here one can again consider regimes of fluid domination, curvature domination of CC domination. Similar to the closed case scenario, if the perfect fluid is dominant for long enough $G_{N}\rho_{0}/a^{3(1+w)} \gg \{t^2_{\Lambda}, a^{-2}\}$, the analysis will resemble the closed universe results with $2$ instead of $\pi$.
Considering island candidates at the time of normalization where $a=a_0$, \eqref{eq:cond_open_enc} reduces to
\be
G_{N} \rho_0>\frac{T_0}{(1+w)2 a_0}~,
\ee
while for backward time evolution we have
\be
 \frac{T_0}{(1+w)2 a_0}\left(\frac{a}{a_0}\right)^2\leq G_{N}\rho_0 \leq \frac{T_0}{(1+w)2 a_0}~.
\ee

Without limiting ourselves to $\chi \gg1$ and $\mathcal{H} \ll 1$, the conditions become the following inequality:
\be
s_c\geq \frac{a^2}{2G_{N}}\times \max\left\{\frac{ 2\sinh^2 \chi}{\sinh 2\chi-2\chi},\,|\mathcal{H}|+|\coth \chi| \right\}~.
\ee
Its fulfillment depends on the exact physical parameters such as the number of degrees of freedom, $c_{th}$, the energy scale $\rho_0$, the entropy density $s_c$, etc.

\section{Recap and Future Directions}
In this work, we have examined when entanglement islands can exist in an FLRW cosmology with a general perfect fluid. We rearranged the necessary conditions for islands derived in Refs.~\cite{Hartman:2020khs, Bousso:2022gth, Espindola:2022fqb} as conditions on the comoving entropy density $s_c$, which is constant in time under the assumption of local thermal equilibrium. In flat universes, we found that time-symmetric slices always have viable island candidates for large enough $\chi$, and these candidates will be islands in the thermofield-doubled model if the reference region is chosen to be the matching region $R$ in $M_R$. For an arbitrary time slice $t_0$ in a flat universe, we found viable island candidates (and thus islands in the thermofield-doubled model) given that $T_0$ and $\rho_0$ are tuned appropriately. Treating the normalization time $t_0$ as an initial time and evolving forwards, we found that even under the assumption that there was no viable island candidate at $t_0$, a flat universe with a perfect fluid develops a viable island candidate provided that $w\geq1$ violating the DEC. Furthermore, $w>1$ implies that the $t_0$ slice will eventually evolve into a slice with a viable island candidate. Evolving backwards in time we found the opposite condition of $|w|<1$ for a viable island candidate and again a condition on the temperature $T_0$ and the energy density $\rho_0$. Finally, we repeated the analysis on closed and open universes with a general perfect fluid. In the presence of spatial curvature, in an expanding universe with monotonically growing $a$, we found that viable island candidates can only exist at the time of normalization or in the past $t\leq t_0$ or up to some finite time after $t_0$, regardless of the equation of state $w$. We applied the conditions to a specific example of a closed universe with a periodic scale factor and no classical singularity, the ``Simple Harmonic Universe" and found viable island candidates on slices near where the scale factor reaches a minimum. Hence, this is a first example where islands exist in a spacetime without a singularity.

Let us now turn to possible future directions. Investigation of the QES prescription is still ongoing, and it is unclear what will be the outcome of its study, in particular in Cosmology. An immediate target is enlarging the numbers of known examples. For example, considering islands in time-dependent cosmologies but with less symmetry, such as anisotropic or inhomogeneous cosmologies. We expect that the spatial anisotropy or inhomogeneity will modify the analysis presented here.

However, such constructions are still toy models. It would be interesting to consider models closer to our own universe, which consists of several fluids with phase transitions and entropy production occurred in the past. It would be valuable to consider a similar analysis with such a thermal history or to apply our analysis to contemporary models of universe, such as the Concordance Model. Furthermore, it would be of interest to investigate whether there could be any observable phenomena related to the existence of islands.

The original island formula involves the generalized entropy and therefore the entanglement entropy of matter. Therefore, another interesting direction is discarding the use of the thermal entropy for a more accurate calculation using the entanglement entropy. Constructing a four dimensional setup with time-dependent background where the entanglement entropy is calculable could yield interesting results and allow us to test the validity of the thermodynamical entropy approximation.

In our analysis and in those of Refs.~\cite{Hartman:2020khs, Bousso:2022gth, Espindola:2022fqb}, it is not clear what the reference spacetime corresponds to in the cosmological setting. It would be preferable to assign a meaningful physical significance to the reference spacetime. For example, one could consider a model involving entanglement between two gravitating universes. Certain attempt have been carried out in that direction~\cite{Balasubramanian:2021wgd}. One of the major motivations of such a development is that it may provide insight on how to apply these results to Multiverse models. The Multiverse consists of multiple universes originating from a parent universe in a never-ending process~\cite{Ben-Dayan:2021ayq,Ijjas:2014nta,Guth:2007ng,Goncharov:1987ir}. Thus, entanglement between two universes seems a reasonable expectation for the Multiverse.

Ref.~\cite{Geng:2021hlu} discussed entanglement wedge reconstruction in the context of islands, arguing that an island must have a coupling with its reference region for it to behave as expected for an entanglement wedge; simple entanglement between the island and the reference system is not enough. This is puzzling, because the thermofield-doubled model used here and in Refs.~\cite{Hartman:2020khs, Bousso:2022gth, Espindola:2022fqb} involve only entanglement, not a coupling. Understanding how to make these results consistent could lead to insight regarding the role of islands in entanglement wedge reconstruction. A first step, for example, could be to perform an analysis similar to ours here and in Refs.~\cite{Hartman:2020khs, Bousso:2022gth, Espindola:2022fqb} using a model with a coupling instead of the thermofield-doubled model.

\paragraph{Acknowledgements} We would like to thank R.~Bousso, T.~Rudelius, and A.~Tajdini for helpful discussions and comments. This work was supported in part by the Berkeley Center for Theoretical Physics; by the Department of Energy, Office of Science, Office of High Energy Physics under QuantISED Award DE-SC0019380 and under contract DE-AC02-05CH11231; and by the National Science Foundation under Award Number 2112880. EW is supported in part by the Berkeley Connect fellowship. IBD was supported in part by the Gale Foundation, RA2100000209.

\bibliographystyle{JHEP}
\bibliography{MIC}

\end{document}